\documentclass[english]{elsarticle}
\usepackage[T1]{fontenc}
\usepackage[latin9]{inputenc}
\usepackage{amsthm}
\usepackage{amsmath}
\usepackage{amssymb}
\usepackage{esint}
\usepackage{color}
\usepackage{epstopdf}

\makeatletter
\theoremstyle{plain}
\newtheorem{thm}{\protect\theoremname}
  \theoremstyle{remark}

\journal{Nuclear Physics B}

\usepackage{babel}
\providecommand{\remarkname}{Remark}
\providecommand{\theoremname}{Theorem}

\makeatother

\usepackage{babel}
  \providecommand{\remarkname}{Remark}
\providecommand{\theoremname}{Theorem}

\begin{document}
\begin{frontmatter}{}

\title{Fermionization, Triangularization and Integrability}

\author[JLU]{Li-Qiang Cai}

\ead{Cailiqiang.m@163.com}

\author[CNU]{Li-Fang Wang}

\ead{Wanglifang@cnu.edu.cn}

\author[ITP,BJUT]{Jian-Feng Wu\corref{cor1}}

\ead{Muchen.Wu@gmail.com}

\author[CNU,BCIIS]{Jie Yang}

\ead{Yang9602@gmail.com}

\author[ITP,SKLTP]{Ming Yu\corref{cor1}}

\ead{Yum@itp.ac.cn}

\cortext[cor1]{Corresponding author}

\address[ITP]{Institute of Theoretical Physics, Chinese Academy of Sciences, Beijing,
China 100190}

\address[SKLTP]{State Key Laboratory of Theoretical Physics, Beijing, China 100190}

\address[BJUT]{Institute of Theoretical Physics, Beijing University of Technology,
Beijing, China 100124}

 \address[JLU]{Department of Mathematics, Jilin University, Changchun, China 130012}

\address[CNU]{School of Mathematics Sciences, Capital Normal University, Beijing,
China 100048}

\address[BCIIS]{Beijing Center for Mathematics and Information Interdisciplinary Sciences,\\ Beijing, China 100048}
\begin{abstract}
In this article, we derive the fermionic formalism of Hamiltonians
as well as corresponding excitation spectrums and states of Calogero-Sutherland(CS),
Laughlin and Halperin systems, respectively. In addition, we study
the triangular property of these Hamiltonians and prove the integrability
in these three cases.\end{abstract}
\end{frontmatter}{}

\section{Introduction and Results\label{sec:Introduction}}

In the area of many body physics, fractional quantum Hall effects
(FQHEs) and integrable models, are two fruitful and important classes.
Many researchers believe these two are connected in a hundred and one
ways. \citep{azuma1994explicit,bergere2000composite,bernevig2008model,estienne2012conformal,ha1994exact,zhang1989effective,zhang1992chern}
Plenty of efforts have been dedicated to find  out the intrinsic relationship.

In FQHEs, the Laughlin trial wavefunction reveals several remarkable
properties of FQHEs at filling number $\nu=\frac{1}{2m+1},$ such
as the fractional statistics as well as the topological orders. Later on
 a conformal field theory (CFT) realization was discovered which shows that the wavefunction
is corresponding to a correlation function of certain vertex operators
. Furthermore this idea is generalized to many other FQH states, e.g. Halperin
state\citep{halperin1984statistics}, Moore-Read state, and Read-Rezayi
state\citep{read1999beyond}, et.al\citep{bergholtz2008quantum,gurarie1997haldane,moore1991nonabelions,wen1994chiral}.
However, CFT is possibly not sufficient to drive the  dynamics of the
edge theory, since it only determines the behavior of the theory near critical
point.\footnote{In the viewpoint of integrable hierarchy, the CFT Hamiltonian $L_0$, is the second Hamiltonian(integral of motion) of the system. However, the finer structures, such as explored in present article and \citep{estienne2012conformal}, are determined by the third or higher level Hamiltonians .}.

A more ambitious thinking is to find the Hamiltonian system behind the
edge ground state. So far, there are two classes of Hamiltonian analysis
for FQHE. One is the Chern-Simons approach, initiated by Zhang, Hansson
and Kivelson in 1989\citep{zhang1989effective}. The other is the
extended Hamiltonian theory, introduced by Murthy and Shankar in late
90's\citep{murthy1999hamiltonian,murthy2003hamiltonian,shankar1997towards}.
The later one contains Chern-Simons as its asymptotic theory.

In our study we try to approach the integrability problem in a different way.
In fact, we are not meant
to establish a unified Hamiltonian theory to solve the complicated
many-body problem.
Instead we are looking for the integrability
behind FQHEs as well as the Hamiltonian expression of it. In order to do
so we separate the excitations of FQHE into two simple
classes: the perturbative class and nonperturbative one.
The nonperturbative class dominates the states in Hilbert space, a.k.a. the basis, the perturbative class organizes those basis into physical states. So perturbations actually
are provided as structure constants (or superposition coefficients). Interestingly, this idea is like
in CFT, where correlation function is made by conformal block and
structure constant (it encodes the multiplicity of the corresponding
conformal block in the correlation function. )  Since the
ground state should not change by perturbations it belongs
to the nonperturbative class. Hence it describes a sort of wave without dissipation which implies that the ground state is a solitonic wave.

 In this way we have related the FQHE theory to soliton theory, the other important area of many-body physics. The question now is to extract excitations from the solitonic wavefunction.
The stable excitations from the soliton wavefunction, are those solutions
of quantum mechanics equation for soliton wave\citep{das1989integrable}.
In this quantum mechanics, the logarithmic of the soliton wavefunction
is a scalar function, while its gradation, gives the effective ``gauge''
potential. Therefore, the Hamiltonian could be written as a Landau-Ginzberg
pseudo-potential form.

Inspired by these observations and a previous work \citep{Wu:2011tt},  we use the same method for Laughlin and Halperin states.
Then we obtain complicated Hamiltonians with non-linear interactions.
However, they are all exact solvable.
 The resolving strategy is as follows: firstly, we interpret the ground state as correlation function in CFT. Secondly, by Jastrow transformation we drop the contribution
of ground state and obtain a relative simple Hamiltonian. Thirdly,
the eigen-equation of the new Hamiltonian can be transformed into
an operator equation acting on the coherent basis. Fourthly,
it turns out that the operator formalism is exactly triangulated. Therefore
we can extract the spectrum as well as the state in a recursive
way. Finally, to analyze the integrability closely, we derive the fermionization
for the bosonic theory. Hence the integrability is clearly
determined by free fermions and the explicit triangularization.

We find, interestingly, the integrability behind Laughlin state, is
the same as the famous Calogero-Sutherland model. Hence the excitations
are those of Jack polynomials\citep{macdonald1995symmetric}. During last two
decades, people claimed that ground states of some FQHEs have the
same properties as those of Jack polynomials. For example the $(k,r,N)$-admissible
representations (labeled by certain restricted Young diagrams) is related
to the filling number $\nu=\frac{k}{r}$ FQHE ground state\citep{azuma1994explicit,bernevig2008generalized,bernevig2008properties,estienne2010clustering,estienne2010electron,estienne2009relating,feigin2002differential,ha1994exact,iso1995collective,lee2014construction}.
 From our viewpoint the basic ingredients are Jack polynomials and additional restrictions, mostly from the fusion rule (which we do not explore in this article),
will rule out some Jack polynomials systematically which results in
the admissible representations.

The Halperin state, corresponding to the two-layer FQHE, shows a secret
integrability dominated by also the triangularization, which says the
number of boxes in Young diagram for the first layer always decreases
while the one for the second layer increases and the total boxes of
these two layers remain the same. The triangularization interaction,
being triple in two kinds of bosonic operators, is quite complicated.
It makes the explicit solution of the excitations slightly difficult.
Nevertheless, since the triangularization is clear, we can give the
explicit solution of the system in principle.

This article is organized as follows. In sec. 2, we review the famous
Calogero-Sutherland model, its operator formalism, the CFT correspondence, the
spectrum and eigenstates. In sec. 3 we obtain the fermionization
of the CS theory followed by the fermionic triangularization and integrability.
In sec. 4 and 5, we provide parallel analysis for Laughlin state
and Halperin state. In sec. 6 we make a conclusion and discuss some further works.

\section{The Calogero-Sutherland Model}

We start our analysis from the famous Calogero-Sutherland(CS) model.
It is an exact solvable model, describing $N$ interacting charged
particles on a unit circle, with two-body interaction
\begin{eqnarray*}
H_{int}=\sum_{i<j} & \frac{\beta(\beta-1)}{\sin^{2}(x_{i}-x_{j})} & \,,
\end{eqnarray*}
in which $x_{i}$ defines the $i$-th particle's position on the circle.
For simplicity, we substitute $\beta=b^{2}$. Then CS Hamiltonian is written as
\begin{equation}
H_{CS}=-\frac{1}{2}\sum_{i=1}^{N}\partial_{i}^{2}+\sum_{i<j}\frac{b^{2}(b^{2}-1)}{\sin^{2}(x_{i}-x_{j})}\,.
\end{equation}

\begin{thm}
\label{theo:isospectrum}$H_{CS}$ is isospectral to another Hamiltonian
\begin{equation}
\tilde{H}_{CS}=-\frac{1}{2}\sum_{i=1}^{N}(\partial_{i}+\partial_{i}\ln\prod_{j<k}\sin^{b^{2}}(x_{j}-x_{k}))(\partial_{i}-\partial_{i}\ln\prod_{r<s}\sin^{b^{2}}(x_{r}-x_{s})\,.
\end{equation}
up to a universal shift of eigen-energy.
\end{thm}
$\!\textbf{Proof of Theorem }$$\ref{theo:isospectrum}$: Defining
the complex coordinate $z_{i}=e^{i2x_{i}}$, we have $\partial_{i}=2iz_{i}\partial_{z_{i}}$.
Therefore
\begin{eqnarray*}
\partial_{i}\ln\prod_{j<k}\sin^{b^{2}}(x_{j}-x_{k}) & = & b^{2}\sum_{\substack { j\\i\neq j}}\cot(x_{i}-x_{j})\\
 & = & ib^{2}\sum_{\substack { j\\i\neq j}}\frac{z_{i}+z_{j}}{z_{i}-z_{j}}\,,
\end{eqnarray*}
and the commutator
\[
\left[\partial_{i}, b^{2}\sum_{\substack{ k \\i\neq k}}\cot(x_{i}-x_{k})\right]=-b^{2}\sum_{\substack{ k \\k\neq i}}\frac{1}{\sin^{2}(x_{k}-x_{i})}.
\]
We can rewrite the $\tilde{H}_{CS}$ as the following formula
\begin{eqnarray*}
\tilde{H}_{CS} & = & -\frac{1}{2}\sum_{i=1}^{N}\partial_{i}^{2}-b^{2}\sum_{i<j}\frac{1}{\sin^{2}(x_{i}-x_{j})}\\
 &  & +\frac{1}{2}b^{4}\sum_{i\neq j,i\neq k}\cot\left(x_{i}-x_{j}\right)\cot\left(x_{i}-x_{k}\right)\,.
\end{eqnarray*}
Using the identity
\begin{eqnarray*}
\sum_{\hbox{distinct \ }i, j, k}\cot\left(x_{i}-x_{j}\right)\cot\left(x_{i}-x_{k}\right)+{i, j, k}\text{{\ cyclic}}\\
=\sum_{\hbox{distinct\ } i, j, k}(-1)=-N(N-1)(N-2) & \,,
\end{eqnarray*}
and the $j=k$ contribution
\[
\sum_{i\neq j}\cot^{2}\left(x_{i}-x_{j}\right)=-\sum_{i\neq j}1+\sum_{i\neq j}\frac{1}{\sin^{2}(x_{i}-x_{j})}\,,
\]
we now have the form of $\tilde{H}_{CS}$ as
\begin{eqnarray}
\tilde{H}_{CS} & = & -\frac{1}{2}\sum_{i=1}^{N}\partial_{i}^{2}+b^{2}(b^{2}-1)\sum_{i<j}\frac{1}{\sin^{2}(x_{i}-x_{j})}-\frac{1}{6}b^4(N-1)N(N+1)\,,\nonumber \\
 & = & H_{CS}-\frac{1}{6}b^4N(N-1)(N+1)
\end{eqnarray}
so Theorem $\ref{theo:isospectrum}$ is proved. Q.E.D.

It is now nature to consider the $\tilde{H}_{CS}$ rather than $H_{CS}$
since the later one, when acting on the ground state, will have a
large energy (proportional to $N^{3}$) contribution to the spectrum.
The ground state of $\tilde{H}_{CS}$ is simply
\begin{equation}
\Psi_{CS}=\prod_{i<j}\sin^{\beta}(x_{i}-x_{j}),\quad \tilde{H}_{CS}\Psi_{CS}=0\text{\,.}\label{eq:GroundCS}
\end{equation}
To extract the spectrums as well as corresponding excitation states,
we need to eliminate the contribution of ground state. It implies
the Jacobi transformation
\[
2H'_{CS}=\Psi_{CS}^{-1}\tilde{H}_{CS}\Psi_{CS}\,.
\]
In this way, we have
\begin{eqnarray}
2H'_{CS} & = & -\frac{1}{2}\sum_{i}(\partial_{i}+2\partial_{i}\ln\Psi_{CS})\partial_{i}\nonumber \\
 & = & -\frac{1}{2}\sum_{i}(2iz_{i}\partial_{z_{i}}+i2b^{2}\sum_{i\neq j}\frac{z_{i}+z_{j}}{z_{i}-z_{j}})(2iz_{i}\partial_{z_{i}})\nonumber \\
 & = & 2\sum_{i}(z_{i}\partial_{z_{i}})^{2}+2b^{2}\sum_{i<j}\frac{z_{i}+z_{j}}{z_{i}-z_{j}}(z_{i}\partial_{z_{i}}-z_{j}\partial_{z_{j}})\,.\label{eq:complexHcs}
\end{eqnarray}

\subsection{Bosonic oscillator formalism of $H'_{CS}$}

The ground state as in ($\ref{eq:GroundCS}$) can be understood as
a CFT correlation function, that is
\[
\Psi_{CS}({z_{i}})\simeq\langle k_{f}|\prod_{i=1}^{N}V_{b}(z_{i})|k_{in}\rangle\,,
\]
with the vertex operator defined by
\[
V_{b}(z)\equiv :e^{b\phi(z)}:\,,
\]
and the bosonic field has the standard mode expansion
\begin{eqnarray*}
\phi(z) & = & q_{0}+p_{0}\ln z+\sum_{n\neq0}\frac{a_{-n}}{n}z^{n}\,,\\
{}[a_{n},a_{m}] & = & n\delta_{n+m,0},\quad [p_{0},q_{0}]=1\,.
\end{eqnarray*}
We can show that briefly. The OPE of vertex operators reads
\[
V_{a}(z)V_{b}(w)=(z-w)^{ab}:V_{a}V_{b}(\frac{z+w}{2}):\,.
\]
If we choose the initial (final) momentum of right (left) vacuum $k_{in}=\frac{b}{2}(1-N)$
($k_{f}=k_{in}+Nb$) , the correlation function is charge neutral and
gives the result
\begin{eqnarray*}
\langle k_{f}|\prod_{i=1}^{N}V_{b}(z_{i})|k_{in}\rangle & = & \prod_{i<j}(z_{i}-z_{j})^{b^{2}}\prod_{i}(z_{i})^{\frac{b^{2}}{2}(1-N)}\\
 & = & \prod_{i<j}\left(\frac{z_{i}-z_{j}}{\sqrt{z_{i}z_{j}}}\right)^{b^{2}}\\
 & = & \prod_{i<j}(2i\sin(x_{i}-x_{j}))^{b^{2}}\,.
\end{eqnarray*}
Therefore up to a constant factor, it is the ground state of CS model. The
excitation state, in principle, will be a state in the Fock space
of the conformal field theory, which in general is a polynomial of
bosonic oscillators. It implies there are one-to-one correspondence
from the excitation wavefunction to an oscillator polynomial. The
basic relation is the coherent relation such that
\begin{equation}
a_{n}\prod_{i}V_{b}^{-}(z_{i})|k_{in}\rangle=b\sum_{i}z_{i}^{n}\prod_{i}V_{b}^{-}(z_{i})|k_{in}\rangle\,.\label{eq:coherent-1}
\end{equation}
It relates the bosonic oscillator mode $a_{m}$to a symmetric polynomial
(or symmetric function if $N\rightarrow\infty$). If we define the
excitation state as follows
\begin{eqnarray*}
\frac{1}{2}\tilde{H}_{CS}\Psi_{\lambda}^{\beta}({z_{i}}) & = & \Psi_{CS}H'_{CS}P_{\lambda}^{\beta}({z_{i}})=E_{\lambda}^{\beta}\Psi_{\lambda}^{\beta}({z_{i}})\\
\Psi_{\lambda}^{\beta}({z_{i}}) & = & \Psi_{CS}({z_{i}})P_{\lambda}^{\beta}({z_{i}})\\
 & \simeq & \langle k_{f}|\prod_{i=1}^{N}V_{b}(z_{i})P_{\lambda}^{\beta}(a^{-})|k_{in}\rangle\,,
\end{eqnarray*}
then
\begin{equation}
\langle k_{f}|P_{\lambda}^{\beta}(a^{+})H'_{CS}(a)\prod_{i=1}^{N}V_{b}^{-}(z_{i})|k_{in}\rangle=E_{\lambda}^{\beta}P_{\lambda}^{\beta}({z_{i}})\,.\label{eq:OpereqCS}
\end{equation}
We have defined here the normal-ordered operator formalism of $H'_{CS}(a)\equiv H$
, such that

\[
H|P_{\lambda}^{\beta}\rangle=E_{\lambda}|P_{\lambda}^{\beta}\rangle.
\]
The next step is to translate the differential formalism Hamiltonian
$H'_{CS}$ into operator formalism with the help of coherent relation
(\ref{eq:coherent-1}). We have the following relations
\begin{eqnarray*}
z_{i}\partial_{z_{i}}\bullet  =  b\sum_{n>0}a_{-n}z_{i}^{n}\bullet,& &
(z_{i}\partial_{z_{i}})^{2}\bullet  =  b^{2}\sum_{n,m>0}a_{-n}a_{-m}z_{i}^{n+m}\bullet
  +b\sum_{n>0}na_{-n}z_{i}^{n}\bullet\\
\sum_{i<j}\frac{z_{i}+z_{j}}{z_{i}-z_{j}}(z_{i}\partial_{z_{i}}-z_{j}\partial_{z_{j}})\bullet & = & b\sum_{i<j,n>0}a_{-n}\frac{z_{i}+z_{j}}{z_{i}-z_{j}}(z_{i}^{n}-z_{j}^{n})\bullet\\
=b\sum_{i<j,n>0}a_{-n}(z_{i}^{n}+2z_{i}^{n-1}z_{j} & + & \cdots2z_{i}z_{j}^{n-1}+z_{j}^{n})\bullet\\
=\left(b\sum_{n,m>0,i,j}a_{-n}z_{i}^{n-m}z_{j}^{m} \right.& + & \left. Nb\sum_{n>0,i=1}^N a_{-n}z_{i}^{n}-b\sum_{n>0,i=1}^N na_{-n}z_{i}^{n}\right)\bullet
\end{eqnarray*} where $\bullet$ denotes $\prod_{j=1}^{N}V_{b}^{-}(z_{j})|k_{i}\rangle$.
So we have the CS Hamiltonian in bosonic operator formalism
\begin{eqnarray}
H=\sum_{n,m>0}b(a_{-n}a_{-m}a_{n+m}+a_{-n-m}a_{n}a_{m})\\
+(1-b^{2})\sum_{n>0}na_{-n}a_{n}+b^{2}N\sum_{n>0}a_{-n}a_{n}\,.\nonumber
\end{eqnarray}
The last term involves the level of corresponding excitations, when
$N\rightarrow\infty$, it overwhelms the excitation spectrum since
it is much larger than other contributions in $H$. In our analysis,
we treat it as the background and we ignore this term. Besides, if we set
\[
\tilde{a}_{-n}=\frac{a_{-n}}{b},\,\,\tilde{a}_{n}=a_{n}b\,,{\text for}\,\,n>0
\]
then we rewrite $H$ as
\begin{eqnarray}
H & = & \sum_{n,m>0}b(b\tilde{a}_{-n}\tilde{a}_{-m}\tilde{a}_{n+m}+\frac{1}{b}\tilde{a}_{-n-m}\tilde{a}_{n}\tilde{a}_{m})\label{deformcs}\\
 & + & (1-b^{2})\sum_{n>0}n\tilde{a}_{-n}\tilde{a}_{n}\nonumber \\
 & = & \sum_{n,m>0}(\tilde{a}_{-n}\tilde{a}_{-m}\tilde{a}_{n+m}+\tilde{a}_{-n-m}\tilde{a}_{n}\tilde{a}_{m})\nonumber \\
 & + & (1-b^{2})\left(\sum_{n>0}n\tilde{a}_{-n}\tilde{a}_{n}-\sum_{n,m>0}\tilde{a}_{-n}\tilde{a}_{-m}\tilde{a}_{n+m}\right)\,.\nonumber
\end{eqnarray}
It is easy to see that  $\tilde{a}$
still holds the Heisenberg algebra so that the fermionization is exact.
 We split the Hamiltonian into free part (the first line of last
equality of (\ref{deformcs})), which is the same as free fermions, and
the interacting part (the second line of last equality of (\ref{deformcs})).

\subsection{Eigenstate and spectrum}

The CS model is exactly solvable. To see that, we first classify the
Fock space expanded by bosons by its level $\mathcal{N}=\sum_{n>0}a_{-n}a_{n}$
such that an arbitrary state
\begin{equation}
|n_{1},n_{2},\cdots,n_{l}\rangle=a_{-n_{1}}a_{-n_{2}}\cdots\, a_{-n_{l}}|0\rangle\,,\label{eq:Fockstate}
\end{equation}
has a level
\[
\mathcal{N}|n_{1},n_{2},\cdots\, n_{l}\rangle=\sum_{i=1}^{l}n_{i}|n_{1},n_{2},\cdots\, n_{l}\rangle\,.
\]
In this classification, there are $P(k)$, the partition number of
$k$, states at a given level $k$. It is easy to check that $H$
commutes with $\mathcal{N}$. Hence they can have common eigenstates.
Therefore, the eigenstate of $H$ can be obtained by a superposition
of states like (\ref{eq:Fockstate}). A closer observation shows that
the Hamiltonian $H$ acting on a state (\ref{eq:Fockstate})
at level $k$ by certain times will definitely generate the lowest
state $|1^{k}\rangle\equiv(a_{-1})^{k}|0\rangle$. We can choose
the coefficients of the Fock state $|1^{k}\rangle$ of all states at level
$k$ to be the same and equal to $b^{-k}$.%
\footnote{
It is a standard choice of a normalized Jack polynomial. While for a non-normalized
Jack polynomial the coefficient can be 1.
}
By this choice, we have removed the irrelevant c-number common factor
of each eigenstate. For example, at level 4, we assume an eigenstate
has the following formalism
\[
|P_{\lambda}^{\beta}\rangle=b^{-4}((a_{-1})^{4}+\alpha_{1}a_{-2}(a_{-1})^{2}+\alpha_{2}a_{-2}^{2}+\alpha_{3}a_{-3}a_{-1}+\alpha_{4}a_{-4})|0\rangle\,,
\]
Thus there are $P(4)-1=5-1=4$ unknown coefficients and also the eigen-energy
$E_{\lambda}^{\beta}$ is not known. However, compare all the coefficients
of the eigen-equation
\[
H|P_{\lambda}^{\beta}\rangle=E_{\lambda}^{\beta}|P_{\lambda}^{\beta}\rangle\,,
\]
we have in total 5 independent equations. They in turn determine
the eigenstate completely. The generalization to level $k$ is then
straightforward.

However, this method does not provide a clear relation between the eigenstate and
the Young diagram underlining the theory. In general, one
can define by hand a sequence of eigen-energies at a given level
so that each state is uniquely related to a Young diagram. But the
reason is weak and unnatural. However, it is quite natural to see
the Young diagram from the fermionic picture, which we will
explore in next section.

\section{Fermionization}

\subsection{Fermionization of free term}
We now rewrite the CS Hamiltonian as $H\equiv H_0 + H_{int}$, here
\begin{equation}\label{eq:freefermion}
H_0 = \sum_{n,m>0}(\tilde{a}_{-n}\tilde{a}_{-m}\tilde{a}_{n+m}+\tilde{a}_{-n-m}\tilde{a}_{n}\tilde{a}_{m})
\end{equation} is the free part, while
\begin{eqnarray}\label{eq:intpart}
H_{int} = (1-b^{2})\left(\sum_{n>0}n\tilde{a}_{-n}\tilde{a}_{n}-\sum_{n,m>0}\tilde{a}_{-n}\tilde{a}_{-m}\tilde{a}_{n+m}\right)
\end{eqnarray} is the interaction part which could be separated into two components
\begin{eqnarray*}H_int &=&H_1+H_2, H_1 = (1-b^{2})(-\sum_{n,m>0}\tilde{a}_{-n}\tilde{a}_{-m}\tilde{a}_{n+m})\\
H_2&=&(1-b^{2})\sum_{n}n\tilde{a}_{-n}\tilde{a}_{n} \end{eqnarray*} for further convenience.
Now we want to fermionize the deformed bosonic Hamiltonian $H$ by
introducing
\begin{equation}
\tilde{a}_{n}=\sum_{r\in\mathbb{Z}+\frac{1}{2}}:\psi_{n-r}\psi_{r}^{*}:\,,
\end{equation}
and also the free Virasoro generator
\begin{eqnarray}
\tilde{T}(z) & = & \frac{1}{2}(\partial_{z}\tilde{\phi}(z))^{2}=-\frac{1}{2}[\psi\partial\psi^{*}+\psi^{*}\partial\psi](z)\\
\tilde{L}_{n} & = & \frac{1}{2}\sum_{m}:\tilde{a}_{n-m}\tilde{a}_{m}:=\sum_{r>0,r\in\mathbb{Z}+\frac{1}{2}}(r+\frac{n}{2}):\psi_{-r}\psi_{n+r}^{*}:
\end{eqnarray}
Firstly, let us consider the free part $H_0$.
Notice that,
\begin{eqnarray}
\sum_{n>0}\tilde{L}_{-n}\tilde{a}_{n} & = & \sum_{n,m>0}\tilde{a}_{-n-m}\tilde{a}_{m}\tilde{a}_{n}+\frac{1}{2}\sum_{n>m>0}\tilde{a}_{-n+m}\tilde{a}_{-m}\tilde{a}_{n}\\
\sum_{n>0}\tilde{a}_{-n}\tilde{L}_{n} & = & \frac{1}{2}\sum_{n>m>0}\tilde{a}_{-n}\tilde{a}_{n-m}\tilde{a}_{m}+\sum_{n,m>0}\tilde{a}_{-n}\tilde{a}_{-m}\tilde{a}_{n+m}\,.
\end{eqnarray}
It gives rise to
\begin{eqnarray}
H_{0}=\sum_{n,m>0}(\tilde{a}_{-n}\tilde{a}_{-m}\tilde{a}_{n+m}+\tilde{a}_{-n-m}\tilde{a}_{n}\tilde{a}_{m})=\frac{2}{3}\left(\sum_{n>0}\tilde{L}_{-n}\tilde{a}_{n}+\tilde{a}_{-n}\tilde{L}_{n}\right)\\
\sum_{n,m>0}\tilde{a}_{-n}\tilde{a}_{-m}\tilde{a}_{n+m}=\frac{2}{3}\left(\sum_{n>0}[2\tilde{a}_{-n}\tilde{L}_{n}-\tilde{L}_{-n}\tilde{a}_{n}]\right)\,.
\end{eqnarray}
Notice that the $H_{0}$ is just the zero mode of the OPE of
\[
\frac{1}{2\pi i}\oint\frac{dz}{z-w}\tilde{T}(z)\partial_{w}\tilde{\phi}(w)\,.
\]
In fermionic representation, we have
\begin{eqnarray}
H_{0}(w) & = & \frac{2}{3}\oint\frac{dz}{2\pi i(z-w)}\left(-\frac{1}{2}[\psi\partial_{z}\psi^{*}+\psi^{*}\partial_{z}\psi])(z)[\psi\psi^{*}](w)\right)\\
 & = & \frac{2}{3}\oint\frac{dz}{2\pi i(z-w)}\left\{ \frac{1}{2}\left(\frac{1}{(z-w)^{3}}+\frac{\psi(z)\psi^{*}(w)}{(z-w)^{2}}-\frac{\partial_{z}\psi^{*}(z)\psi(w)}{z-w}\right)\right.\nonumber \\
 &  & +\left.\frac{1}{2}\left(-\frac{1}{(z-w)^{3}}-\frac{\psi^{*}(z)\psi(w)}{(z-w)^{2}}+\frac{\partial_{z}\psi(z)\psi^{*}(w)}{z-w}\right)\right\} \nonumber \\
 & = & \frac{1}{3}\oint\frac{dz}{2\pi i(z-w)}\left\{ \frac{\psi(z)\psi^{*}(w)-\psi^{*}(z)\psi(w)}{(z-w)^{2}}\right.\nonumber \\
 &  & +\left.\left(\frac{\partial_{z}\psi(z)\psi^{*}(w)-\partial_{z}\psi^{*}(z)\psi(w)}{z-w}\right)\right\} \nonumber \\
 & = & \frac{1}{2}\left([(\partial_{w})^{2}\psi(w)]\psi^{*}(w)-[(\partial_{w})^{2}\psi^{*}(w)]\psi(w)\right)\,.\nonumber
\end{eqnarray}
The operator formalism $H_{0}$ is
\begin{eqnarray}
H_{0} & = & \frac{1}{2\pi i}\oint w^{2}dwH_{0}(w)\\
 & = & \frac{1}{2}\left([(\partial_{w})^{2}\psi(w)]\psi^{*}(w)-[(\partial_{w})^{2}\psi^{*}(w)]\psi(w)\right)\nonumber \\
 & = & \frac{1}{2}\sum_{r\in\mathbb{Z}+\frac{1}{2}}:\psi_{-r}\psi_{r}^{*}:\left((-r-\frac{1}{2})(-r-\frac{3}{2})+(r-\frac{1}{2})(r-\frac{3}{2})\right)\nonumber \\
 & = & \sum_{r>0,r\in\mathbb{Z}+\frac{1}{2}}\left(r^{2}+\frac{3}{4}\right)(\psi_{-r}\psi_{r}^{*}-\psi_{-r}^{*}\psi_{r})\,.\nonumber
\end{eqnarray}

\subsection{Fermionization of triple bosons term}

Now let us consider the term
\begin{eqnarray*}
 &  & (1-b^{2})(-\sum_{n,m>0}\tilde{a}_{-n}\tilde{a}_{-m}\tilde{a}_{n+m})\\
 &  & =\frac{2}{3}(1-b^{2})\sum_{n}[-2\tilde{a}_{-n}\tilde{L}_{n}+\tilde{L}_{-n}\tilde{a}_{n}]\\
 &  & =\frac{2}{3}(1-b^{2})\left(\sum_{n}{{\diamond}\atop {\diamond}}\tilde{L}_{-n}\tilde{a}_{n}-2\tilde{a}_{-n}\tilde{L}_{n}{{\diamond}\atop {\diamond}}+{\text{Contractions}}\right)\,.
\end{eqnarray*}
In formulation of $\tilde{a}$ and $\tilde{L}$, it is easier to calculate
the fermionic expression. We have
\begin{eqnarray}
{{\diamond}\atop {\diamond}}\tilde{L}_{-n}\tilde{a}_{n}-2\tilde{a}_{-n}\tilde{L}_{n}{{\diamond}\atop {\diamond}} & = & \sum_{\begin{subarray}{c}
r\in\mathbb{Z}+\frac{1}{2}\\
s\in\mathbb{Z}+\frac{1}{2}
\end{subarray}}\left((r-\frac{n}{2}):\psi_{-r}\psi_{-n+r}^{*}\psi_{-s}\psi_{n+s}^{*}:\right.\\
 &  & -\left.(2s+n):\psi_{-r}\psi_{-n+r}^{*}\psi_{-s}\psi_{n+s}:\right)\nonumber \\
 & = & \sum_{\begin{subarray}{c}
r,s,k,l\in\mathbb{Z}+\frac{1}{2}\\
r+s<0,l+k>0\\
r+s+k+l=0
\end{subarray}}:\psi_{r}\psi_{s}^{*}\psi_{k}\psi_{l}^{*}:\left(\frac{1}{2}(s-r)+k-l\right)\,.\nonumber
\end{eqnarray}
To calculate the contractions, we need to calculate the commutator
\begin{eqnarray}
[\tilde{L}_{-n},\psi_{-k}] & = & \sum_{r}\left[(r-\frac{n}{2}):\psi_{-r}\psi_{-n+r}^{*}:\,,\,\,\psi_{-k}\right]\\
 & = & \left(\theta(k>0)\sum_{r>0}\delta_{r-n,k}\psi_{-r}\right.\nonumber \\
 &  & +\left.\theta(k<0)\sum_{r<0}\delta_{r-n,k}\psi_{-r}\right)(r-\frac{n}{2})\nonumber \\
 & = & (k+\frac{n}{2})\psi_{-n-k}\nonumber \\
\,[\tilde{L}_{-n},\psi_{n+k}^{*}] & = & \sum_{r}\left[(r-\frac{n}{2}):\psi_{-r}\psi_{-n+r}^{*}:\,,\,\,\psi_{n+k}^{*}\right]\\
 & = & \left(-\theta(n+k>0)\sum_{r>0}\delta_{r,n+k}\psi_{-n+r}^{*}\right.\nonumber \\
 &  & -\left.\theta(n+k<0)\sum_{r<0}\delta_{r,n+k}\psi_{-n+r}^{*}\right)(r-\frac{n}{2})\nonumber \\
 & = & -(k+\frac{n}{2})\psi_{k}^{*}\nonumber \\
\,[\psi_{-k},\tilde{L}_{n}] & = & (-k+\frac{n}{2})\psi_{n-k}\\
\,[\psi_{-n+k}^{*},\tilde{L}_{n}] & = & (k-\frac{n}{2})\psi_{k}^{*}\,.
\end{eqnarray}
Hence we have the contraction as follows
\begin{eqnarray}
{\text{Contractions}} & = & \left(\sum_{n\in\mathbb{Z},k>0}:[\tilde{L}_{-n},\psi_{-k}]\psi_{n+k}^{*}:+\sum_{n\in\mathbb{Z},n+k<0}:\psi_{-k}[\tilde{L}_{-n},\,\psi_{n+k}^{*}]:\right.\nonumber \\
 &  & -2\left.\left(\sum_{n\in\mathbb{Z},n-k<0}:\psi_{-k}[\psi_{-n+k}^{*},\tilde{L}_{n}]:+\sum_{n\in\mathbb{Z},k<0}:[\psi_{-k},\,\tilde{L}_{n}]\psi_{-n+k}^{*}:\right)\right)\nonumber \\
 & = & \left(\sum_{k>0}(k+\frac{n}{2}):\psi_{-n-k}\psi_{n+k}^{*}:-\sum_{n+k<0}(k+\frac{n}{2}):\psi_{-k}\psi_{k}^{*}:\right.\nonumber \\
 &  & -2\left.\sum_{n<k}(k-\frac{n}{2}):\psi_{-k}\psi_{k}^{*}:+2\sum_{k<0}(k-\frac{n}{2}):\psi_{n-k}\psi_{-n+k}^{*}:\right)\nonumber \\
 & = & \sum_{k>n}(k-\frac{n}{2}):\psi_{-k}\psi_{k}:-\sum_{k<-n}(k+\frac{n}{2}):\psi_{-k}\psi_{k}^{*}:\nonumber \\
 &  & +\sum_{k<-n}(2k+n):\psi_{-k}\psi_{k}^{*}:-\sum_{k>n}(2k-n):\psi_{-k}\psi_{k}^{*}:\nonumber \\
 & = & \left(\sum_{k>n}(\frac{n}{2}-k)+\sum_{k<-n}(k+\frac{n}{2})\right):\psi_{-k}\psi_{k}^{*}:\nonumber \\
 & = & \sum_{0<n<k}(\frac{n}{2}-k)(\psi_{-k}\psi_{k}^{*}-\psi_{-k}^{*}\psi_{k})\\
 & = & \sum_{k>0}\left((-k)(k-\frac{1}{2})+\frac{1}{4}(k+\frac{1}{2})(k-\frac{1}{2})\right)(\psi_{-k}\psi_{k}^{*}-\psi_{-k}^{*}\psi_{k})\nonumber \\
 & = & -\frac{1}{16}\sum_{k>0,k\in\mathbb{Z}+\frac{1}{2}}(6k-1)(2k-1)(\psi_{-k}\psi_{k}^{*}-\psi_{-k}^{*}\psi_{k})
\end{eqnarray}
Now the whole $H_{1}$ is
\begin{eqnarray}
H_{1} & = & \frac{2}{3}(1-b^{2})\left(\sum_{\begin{subarray}{c}
r,s,k,l\in\mathbb{Z}+\frac{1}{2}\\
r+s<0,l+k>0\\
r+s+k+l=0
\end{subarray}}:\psi_{r}\psi_{s}^{*}\psi_{k}\psi_{l}^{*}:\left(\frac{1}{2}(s-r)+k-l\right)\right.\\
 &  & -\left.\frac{1}{16}\sum_{k>0,k\in\mathbb{Z}+\frac{1}{2}}(6k-1)(2k-1)(\psi_{-k}\psi_{k}^{*}-\psi_{-k}^{*}\psi_{k})\right)\nonumber
\end{eqnarray}

\subsection{Fermionization of double bosons term}

The last term in the Hamiltonian is the term
\[
H_{2}=(1-b^{2})\sum_{n}n\tilde{a}_{-n}\tilde{a}_{n}\,,
\]
fermionization leads to an expression
\begin{eqnarray}
H_{2} & = & (1-b^{2})\left(\sum_{r,k\in\mathbb{Z}+\frac{1}{2}}n:\psi_{k}\psi_{-n-k}^{*}\psi_{r}\psi_{n-r}^{*}:+\text{Contractions}\right)\label{H2}\\
 & = & (1-b^{2})\left(\sum_{\begin{subarray}{c}
r+s<0,k+l>0\\
r+s+k+l=0\\
r,s,k,l\in\mathbb{Z}+\frac{1}{2}
\end{subarray}}(k+l):\psi_{r}\psi_{s}^{*}\psi_{k}\psi_{l}^{*}:+\text{Contractions}\right)\,.\nonumber
\end{eqnarray}
The contractions in (\ref{H2}) now can be calculated as
\begin{eqnarray}
\text{Contractions} & = & \sum_{\begin{subarray}{c}
n>0\\
k,l>0
\end{subarray}}n\left((\psi_{-k}\psi_{k-n}^{*}-\psi_{-n-k}^{*}\psi_{k})(\psi_{-l}\psi_{n+l}^{*}-\psi_{n-l}^{*}\psi_{l})\right)\nonumber \\
 &  & -\sum_{r,k\in\mathbb{Z}+\frac{1}{2}}n:\psi_{k}\psi_{-n-k}^{*}\psi_{r}\psi_{n-r}^{*}:\nonumber \\
 & = & \sum_{k>l>0}(k-l)\psi_{-k}\psi_{k}^{*}+\sum_{0<k<l}(l-k)\psi_{-l}^{*}\psi_{l}\nonumber \\
 & = & \sum_{k>0}(\frac{(k-\frac{1}{2})(k+\frac{1}{2})}{2})(\psi_{-k}\psi_{k}^{*}+\psi_{-k}^{*}\psi_{k})\nonumber \\
 & = & \sum_{k>0}\frac{1}{2}(k^{2}-\frac{1}{4})(\psi_{-k}\psi_{k}^{*}+\psi_{-k}^{*}\psi_{k})\,.
\end{eqnarray}

\subsection{Full expression}

Combining all the expressions, we obtain the total $H$
\begin{eqnarray}\label{eq:full_Hamiltonian}
H & = & H_{0}+H_{1}+H_{2}=\sum_{k>0,k\in\mathbb{Z}+\frac{1}{2}}\left(k^{2}+\frac{3}{4}\right)(\psi_{-k}\psi_{k}^{*}-\psi_{-k}^{*}\psi_{k})\\
 &  & +\frac{2}{3}(1-b^{2})\left(\sum_{\begin{subarray}{c}
r,s,k,l\in\mathbb{Z}+\frac{1}{2}\\
r+s<0,l+k>0\\
r+s+k+l=0
\end{subarray}}:\psi_{r}\psi_{s}^{*}\psi_{k}\psi_{l}^{*}:\left(\frac{1}{2}(s-r)+k-l\right)\right.\nonumber \\
 &  & -\left.\frac{1}{16}\sum_{k>0,k\in\mathbb{Z}+\frac{1}{2}}(6k-1)(2k-1)(\psi_{-k}\psi_{k}^{*}-\psi_{-k}^{*}\psi_{k})\right)\nonumber \\
 &  & +(1-b^{2})\left(\sum_{\begin{subarray}{c}
r+s<0,k+l>0\\
r+s+k+l=0\\
r,s,k,l\in\mathbb{Z}+\frac{1}{2}
\end{subarray}}(k+l):\psi_{r}\psi_{s}^{*}\psi_{k}\psi_{l}^{*}:\right.\nonumber \\
 &  & +\left.\sum_{k>0}\frac{1}{2}(k^{2}-\frac{1}{4})(\psi_{-k}\psi_{k}^{*}+\psi_{-k}^{*}\psi_{k})\right)\nonumber \\
 & = & H_{0}+(1-b^{2})H_{d}+(1-b^{2})H_{t}\,.\nonumber
\end{eqnarray}
Here we introduce
\begin{eqnarray}
H_{d} & = & \sum_{k>0,k\in\mathbb{Z}+\frac{1}{2}}\frac{1}{3}(k-\frac{1}{2})\psi_{-k}\psi_{k}^{*}+(k-\frac{1}{2})(k+\frac{1}{6})\psi_{-k}^{*}\psi_{k}\\
 &  & +\sum_{\begin{subarray}{c}
k+l>0\\
k,l\in\mathbb{Z}+\frac{1}{2}
\end{subarray}}\frac{2}{3}(2k+l):\psi_{-l}\psi_{-k}^{*}\psi_{k}\psi_{l}^{*}:\,,\nonumber \\
H_{t} & = & \sum_{\begin{subarray}{c}
k+l>0\\
s,k,l\in\mathbb{Z}+\frac{1}{2}
\end{subarray}}(2k+\frac{2}{3}(s+l)):\psi_{-s-k-l}\psi_{s}^{*}\psi_{k}\psi_{l}^{*}:
\end{eqnarray}

\subsection{Fermionic Triangularization}

\subsubsection{$H_{d}$ shifts eigen-energy}

Now it is clear that $$H_0^\beta\equiv H_{0}+(1-b^{2})H_{d}$$ preserves the Schur state,
but $H_{d}$ changes the eigen-energy. The aim here is  to prove
the eigen-energy
\begin{equation}
E_{\lambda}^{\beta}=E_{\lambda}^{1}+(1-b^{2})\sum_{i}^{\ell(\lambda^{t})}(\lambda_{i}^{t})^{2}\, ,
\end{equation}
where
\begin{equation}
E_{\lambda}^{1}=\sum_{i}^{d(\lambda)}(\lambda_{i}-i+\frac{1}{2})^{2}-(\lambda_{i}^{t}-i+\frac{1}{2})^{2}
\end{equation}
is the eigen-energy of free fermions excitation labeled by Young diagram
$\lambda$. For simplicity, we define
\begin{eqnarray}
n_{i}(\lambda)=\lambda_{i}-i+\frac{1}{2},\,\, m_{i}(\lambda)=\lambda_{i}^{t}-i+\frac{1}{2}\,.
\end{eqnarray}

\begin{thm}
The eigen-energy of $H_{d}$ related to Schur state $\lambda$ is
\[
E_{\lambda}^{d}=\sum_{i}^{\lambda_{1}}(\lambda_{i}^{t})^{2}\,.
\]
\label{Theo:Hdenergy}
\end{thm}
\begin{figure}
  \centering
  \includegraphics[width=.2\textwidth]{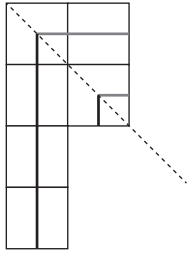}\\
  \caption{The (2,2,1,1) Young diagram}\label{fig:Fig1}
\end{figure}

Before proving this theorem, we now consider an example $\lambda=\{2,2,1,1\}$
as shown in Fig. \ref{fig:Fig1}. The corresponding Schur state is
\[
|\lambda\rangle=(-)\psi_{-3/2}\psi_{-7/2}^{*}\psi_{-1/2}\psi_{-1/2}^{*}|vac\rangle\,.
\]
From the formalism of $H_{d}$, the energy eigenvalue is contributed
by the following terms
\begin{eqnarray*}
 &  & \frac{2}{3}\left(\frac{3}{2}-1\right)\psi_{-3/2}\psi_{3/2}^{*}\,,\,\left((\frac{7}{2}-\frac{1}{2})(\frac{7}{2}+\frac{1}{6})\right)\psi_{-7/2}^{*}\psi_{7/2}\\
 &  & \frac{2}{3}\left\{ \left(7+\frac{3}{2}\right)\psi_{-3/2}\psi_{-7/2}^{*}\psi_{7/2}\psi_{3/2}^{*}+\left(1+\frac{1}{2}\right)\psi_{-1/2}\psi_{-1/2}^{*}\psi_{1/2}\psi_{-1/2}^{*}\right.\\
 &  & \hspace{0.5cm}+\left(7+\frac{1}{2}\right)\psi_{-1/2}\psi_{-7/2}^{*}\psi_{7/2}\psi_{1/2}^{*}+\left(1+\frac{3}{2}\right)\psi_{-3/2}\psi_{-1/2}^{*}\psi_{1/2}\psi_{3/2}^{*}\\
 &  & \left.\hspace{0.5cm}+\left(1-\frac{3}{2}\right)\psi_{-3/2}\psi_{-1/2}\psi_{-1/2}^{*}\psi_{3/2}^{*}-\left(7-\frac{1}{2}\right)\psi_{-1/2}^{*}\psi_{-7/2}^{*}\psi_{7/2}\psi_{1/2}\right\} \,.
\end{eqnarray*}
Combine all of them, we have the eigenvalue of $H_{d}$
\[
E_{\lambda}^{d}=\frac{1}{3}+11+\frac{2}{3}(10+10-7)=20=16+4\,.
\]
{\textbf{ Proof of Theorem \ref{Theo:Hdenergy}}}

\hspace{-0.5cm}For a generic Schur state, the proof of Theorem \ref{Theo:Hdenergy}
is  following. A generic Schur state is denoted by
\begin{equation}
|\lambda\rangle=(-)^{\sum_{i}^{d(\lambda)}\left(m_{i}-\frac{1}{2}\right)}\prod_{i}^{d(\lambda)}\psi_{-n_{i}}\psi_{-m_{i}}^{*}|vac\rangle\,.
\end{equation}
Acting on it, the first two terms of $H_{d}$ contribute
\begin{eqnarray}
\sum_{i}^{d(\lambda)}\frac{1}{3}(n_{i}-\frac{1}{2})+\left(m_{i}-\frac{1}{2}\right)\left(m_{i}+\frac{1}{6}\right)\,.
\end{eqnarray}
The third term  contains
\[
\left(\begin{array}{c}
2d(\lambda)\\
2
\end{array}\right)=d(\lambda)(2d(\lambda)-1)
\]
terms. We can have three independent picking strategies. Among them
there are $d(\lambda)^{2}$ terms picked from  a pair of
$\psi$ and $\psi^{*}$ which we call $\psi\psi^{*}$-strategy.
Other terms are from two fermionic modes picked from either the set of all $\psi$'s
or $\psi^{*}$'s (the $\psi\psi$-strategy and $\psi^{*}\psi^{*}$-strategy).

We first consider the $\psi\psi^{*}$ -strategy. The contribution is
\begin{equation}
E_{\psi\psi^{*}}=\frac{2}{3}\sum_{i,j=1}^{d(\lambda)}(2m_{i}+n_{j})\,.
\end{equation}
The $\psi\psi$ strategy has a signature contribution (-1). Its
contribution to the total energy is
\begin{equation}
E_{\psi\psi}=\frac{2}{3}\sum_{i>j}^{d(\lambda)}(2n_{i}-n_{j})\,.
\end{equation}
Similarly, the $\psi^{*}\psi^{*}$ -strategy gives rise to
\begin{equation}
E_{\psi^{*}\psi^{*}}=\frac{2}{3}\sum_{i<j}^{d(\lambda)}(-2m_{i}+m_{j})\,.
\end{equation}
The summation conditions $i>j$ or $i<j$ reflects the condition $k+l>0$.
Thus we have translated the Theorem \ref{Theo:Hdenergy} into the form
that
\begin{eqnarray}
 &  & \sum_{i}^{\ell(\lambda)}(\lambda_{i}^{t})^{2}=\sum_{i}^{d(\lambda)}\frac{1}{3}(n_{i}-\frac{1}{2})+\left(m_{i}-\frac{1}{2}\right)\left(m_{i}+\frac{1}{6}\right)\label{imptcombine}\\
 &  & +\frac{2}{3}\left(\sum_{i,j=1}^{d(\lambda)}(2m_{i}+n_{j})+\sum_{i>j}^{d(\lambda)}(2n_{i}-n_{j})+\sum_{i<j}^{d(\lambda)}(-2m_{i}+m_{j})\right)\,.\nonumber
\end{eqnarray}
We use mathematical induction to prove it here. A direct proof can be found in Appendix.

Suppose the relation holds for a Young diagram $\lambda$,
with $|\lambda|=n$. We will prove it holds for adding one more box to $\lambda$
under or right to the certain position $(i,j)$. These two different cases are following.

\hspace{-0.5cm}\textbf{Case I}: $\lambda_{j}^{t}\geq j$ $ \lambda^t_{j-1}> i$
 and the box is attached just under the $(i,j)$ box, so that
\[
m_{j}\rightarrow m_{j}+1\,.
\]

The change of R.H.S of (\ref{imptcombine}) will be
\begin{eqnarray}
\Delta_{RHS} & = & \left(2m_{j}+\frac{2}{3}\right)+\frac{2}{3}\left(2d(\lambda)+j-1-2(d(\lambda)-j)\right)\\
 & = & 2m_{j}+2j=2\lambda_{j}^{t}+1=\Delta_{LHS}\,.\nonumber
\end{eqnarray}

\hspace{-0.5cm}\textbf{Case II}: $\lambda_{i}\geq i$  $\lambda_{i-1}>j$,  the box is
attached to the right of the box $(i,j)$, so that
\[
n_{i}\rightarrow n_{i}+1\,.
\]

The change of R.H.S of (\ref{imptcombine}) will be
\begin{eqnarray}
\Delta_{RHS} & = & \frac{1}{3}+\frac{2}{3}(3i-2)=2i-1\,.
\end{eqnarray}
Notice that the $j+1$-th column has $\lambda_{j+1}^{t}=i-1$, so
the change of L.H.S of (\ref{imptcombine}) is
\[
\Delta_{L.H.S}=i^{2}-(i-1)^{2}=2i-1=\Delta_{R.H.S}\,.
\]
Now we conclude that the identity (\ref{imptcombine}) holds for any
Young diagram and the Theorem \ref{Theo:Hdenergy} is proved. Q.E.D.

\subsubsection{$H_{t}$ squeezes the state}

As argued previously, the $H_t$ always squeezes the
original Young diagram $\lambda$ for a given Schur state to some ``thinner''
 $\lambda^{\prime}$'s. To be precise  after its action
\begin{equation}
\lambda^{\prime}<\lambda \quad \hbox{meaning} \quad \sum_{i=1}^{j}\lambda_{i}^{\prime}<\sum_{i=1}^{j}\lambda_{i},\,\text{for }\, j=1,2,\cdots\,.\label{squeeze}
\end{equation}
To see this triangular property in a more transparent form, we first
reduce the summation area from $r+s<0,k+l>0$ to  $k+l>0,k+s<0,k>r$ and $k<r$.
The $k<r$ case can be obtained from re-labeling the index ($r\leftrightarrow k$)
and exchanging $\psi_{r}$ and $\psi_{k}$
\begin{eqnarray}
\frac{1}{1-b^{2}}H_{t} & = & \sum_{\begin{subarray}{c}
k+l>0, r+s+k+l=0\\
s,k,l\in\mathbb{Z}+\frac{1}{2}
\end{subarray}}\left(\frac{2}{3}(2k-r)\right):\psi_{r}\psi_{s}^{*}\psi_{k}\psi_{l}^{*}:\\
 & = & \sum_{\begin{subarray}{c}
k+l>0,k+s<0\\
k>r,r+s+k+l=0
\end{subarray}}\left(\frac{2}{3}(2k-r-2r+k)\right):\psi_{r}\psi_{s}^{*}\psi_{k}\psi_{l}^{*}:\\
 & = & \sum_{\begin{subarray}{c}
k+l>0,k+s<0\\
k>r,r+s+k+l=0
\end{subarray}}2(k-r):\psi_{r}\psi_{s}^{*}\psi_{k}\psi_{l}^{*}: \, .
\end{eqnarray}
$H_{t}$ is simplified and the summation area is decomposed into five cases
\begin{eqnarray}
\frac{1}{1-b^{2}}H_{t} & = & 2\sum_{\begin{subarray}{c}
k+l>0,k+s<0\\
k>r,k+l+r+s=0
\end{subarray}}(k-r):\psi_{r}\psi_{s}^{*}\psi_{k}\psi_{l}^{*}:\label{eq:triangularH}\\
 & = & 2\sum_{n=1,k>r>0}^{n=s-1/2}(k-r)\psi_{-k-n}^{*}\psi_{-r+n}^{*}\psi_{r}\psi_{k}\nonumber \\
 &  & +2\sum_{n=1,k>r>0}^{n=[(k-r-1)/2]}(r-k+2n)\psi_{-k+n}\psi_{-r-n}\psi_{r}^{*}\psi_{k}^{*}\nonumber \\
 &  & +2\sum_{n=1,k,r>0}^{n=r-1/2}(k+r-n)\psi_{-r+n}\psi_{-k-n}^{*}\psi_{k}\psi_{r}^{*}\\
 &  & +2\sum_{k>r>0,s>0}(r-k)\psi_{-r-s-k}^{*}\psi_{r}\psi_{k}\psi_{s}^{*}\nonumber \\
 &  & +2\sum_{k>r>0,s>0}(k-r)\psi_{-k}\psi_{-s}^{*}\psi_{-r}\psi_{r+s+k}^{*}\,.\nonumber
\end{eqnarray}
Acting on a Schur state corresponding to a Young diagram $H_t$ gives rise to the following five processes.

\hspace{-0.5cm}\textbf{Case I:}The first line of the second equality
in (\ref{eq:triangularH}) annihilates two columns of the corresponding
Young diagram of length $r$ and $k$ (k>r) and creates two new
columns of length $k+n$ and $r-n$, which makes long column longer
and short column shorter simultaneously. So it squeezes the Young
diagram.

\hspace{-0.5cm}\textbf{Case II:} The second line annihilates two
rows of length $r$ and $k$ (k>r) and generates two new rows of
length $k-n$ and $r+n$, which makes long row shorter and short row
longer, but not longer than the new shorter row, also, it squeezes the Young diagram.

\hspace{-0.5cm}\textbf{Case III:} The third line annihilates one
row of length $k$ and one column $r$ and generates a shorter row (length
$r-n$) and a longer column (length $k+n$).

\hspace{-0.5cm}\textbf{Case IV:} The fourth line annihilates two
column of length $r$ and $k$ (k>r) and one row of length $s$
and generates a single column of length $r+k+s$ .

\hspace{-0.5cm}\textbf{Case V:} The fifth line annihilates a long
row of length $r+k+s$ into three short columns of lengths $r$,
$k$, $s$ respectively .

From the above analysis, we conclude that the $H_{t}$, when acting
on a Schur state, will generate series of squeezed states. We call
this property the fermionic triangularization.

\subsection{Integrability}

The triangular property means the eigenstate can be understood as
$|P_{\lambda}(\psi\psi^{*})\rangle$, with
\[
P_{\lambda}=s_{\lambda}+\sum_{\mu<\lambda}c_{\lambda,\mu}s_{\mu}\,.
\]
This ansatz is due to the eigenstate of bosonic $H$ is a function
of bosonic oscillators $a_{-n}$'s. When $a_{-n}$ acts on a coherent basis,
the eigenstate will be a power sum symmetric function
\begin{eqnarray}
 &  & \tilde{a}_{-n}\exp\left(\sum_{i,n}b\frac{a_{n}}{-n}z_{i}^{n}\right)|0\rangle=\exp\left(\sum_{i,n}b\frac{a_{n}}{-n}z_{i}^{n}\right)\sum_{i}z_{i}^{n}|0\rangle\nonumber \\
 &  & \Rightarrow\tilde{a}_{-n}\simeq\sum_{i}z_{i}^{n}=p_{n}(z_{i})\,.
\end{eqnarray}
It in turn determines the eigenstate itself as a symmetric function.
The eigenvalue of $H$ is
\begin{eqnarray}
H|P_{\lambda}(\psi\psi^{*})\rangle=E_{\lambda}^{\beta}|P_{\lambda}(\psi\psi^{*})\rangle\,.\label{eigeneqn}
\end{eqnarray}
Let us consider the leading Schur function $s_{\lambda}$. Because $H_{t}$ changes basis,
the eigenvalue should come from  the action of $H_{0}^{\beta}$
on
\[
|s_{\lambda}\rangle=s_{\lambda}(\psi\psi^{*})|vac\rangle\,.
\]

The equation (\ref{eigeneqn}), has an explicit solution (up to a constant similarity transformation) that
\begin{equation*}
|P_{\lambda}\rangle \propto  R(E)|s_{\lambda}\rangle\,,
\end{equation*}
where
\begin{equation}
R(E)  =  \frac{1}{1-\frac{1-b^{2}}{E_{\lambda}^{\beta}-H_{0}^{\beta}}H_{t}}\,.\label{explicitsol}
\end{equation}
This solution can be derived as follows. We first rewrite $H$
as
\begin{equation}
H=E-(E-H_{0}^{\beta})\left(1-\frac{1-b^{2}}{E-H_{0}^{\beta}}H_{t}\right)\,\label{decompotrick}
\end{equation}
then
\begin{eqnarray}
H|P_{\lambda}\rangle & = & \left(E_{\lambda}^{\beta}-(E_{\lambda}^{\beta}-H_{0}^{\beta})\left(1-\frac{1-b^{2}}{E_{\lambda}^{\beta}-H_{0}^{\beta}}H_{t}\right)\right)R(E)|s_{\lambda}\rangle\\
 & = & E_{\lambda}^{\beta}R(E)|s_{\lambda}\rangle-(E_{\lambda}^{\beta}-H_{0}^{\beta})|s_{\lambda}\rangle=E_{\lambda}^{\beta}|P_{\lambda}\rangle\,.\nonumber
\end{eqnarray}

Notice that till now we have used the deformed bosonic oscillators
$\tilde{a}$. It is not convenient when we consider the standard symmetric
function formulae. We need to introduce a similarity transformation
that transforms $\tilde{a}$'s back to $a$'s. \\

\begin{equation}
D=\exp\bigl(-\log(b)(\tilde{q}\tilde{a}_{0}+\sum_{n>0}\tilde{a}_{-n}\tilde{a}_{n}/n)\bigr)\,,
\end{equation}
or equivalently, we have a fermionic formalism
\begin{equation}
D_{\pm}=b^{\pm\frac{1}{2}\sum_{r>0}(\psi_{-r}\psi_{r}^{*}+\psi_{-r}^{*}\psi_{r})}\,,
\end{equation}
where $\pm$ means acting on bra(left) or ket(right) state respectively.

\section{Laughlin state and its Hamiltonian}

The Laughlin state is defined as
\begin{eqnarray}
\Psi_{L}(\{z_{i}\})=\prod_{i<j}(z_{i}-z_{j})^{b^{2}}\exp\left(-\sum_{i}\frac{|z_{i}^{2}|}{4\ell}\right)\,,
\end{eqnarray}
where  $b^{-2}=\nu\in\mathbb{Z}$ is the filling fraction. In the formula
\begin{equation*}
\ell=\sqrt{\frac{\hbar}{eB}}
\end{equation*}
is the basic magnetic length scale. Usually  it is normalized to be 1. $e$ stands for
the electron charge and $B$ is the external magnetic field strength.
The Gaussian factor
\[
\exp\left(-\sum_{i}\frac{|z_{i}^{2}|}{4\ell}\right)\,,
\]
will be ignored later on since it will bring in excessive clutters.

The Laughlin wavefunction can be understood as follows. Consider
there are $N$ quasi-particles containing in the interior of a disk
of area $A_{N}=2N\pi b^{2}$. The ground state of this system is
 a correlation function of these $N$ free quasi-particles in a
background magnetic field, which in general makes the total charge
to be zero, that is, a neutral correlation function. We define the
vertex operator for a quasi-particle
\begin{eqnarray*}
 &  & V_{b}(z)=e^{b\phi(z)}\,.\,
\end{eqnarray*}
The density $\rho_{0}$ of the $\phi$ field on the disk, for a ground
state, should be uniform anywhere. Otherwise a density wave will be
excited. It is simply
\begin{eqnarray*}
\rho_{0} & = & \frac{N}{2\pi Nb^{2}}=\frac{1}{2\pi b^{2}}\,.
\end{eqnarray*}
Thus the background charge is
\[
S=e^{-b\int d^{2}w\rho_{0}\phi(w)}\,.
\]
Without this factor, the correlation function will vanish. Now the correlation
function is written as
\begin{eqnarray}
\langle0\prod_{i=1}^{N}V_{b}(z_{i})S|0\rangle & = & \Bigg\langle\prod_{i=1}^{N}e^{b\phi(z)}\exp\bigg(-b\int d^{2}w\rho_{0}\phi(w)\bigg)\Bigg\rangle\nonumber \\
 & = & \prod_{i<j}^{N}(z_{i}-z_{j})^{b^{2}}\equiv\tilde{\Psi}_{L}(\{z_{i}\})
\end{eqnarray}
Here we have defined the so-called \textquotedbl{}reduced wave function\textquotedbl{}
$\tilde{\Psi}_{L}(\{z_{i}\})$ without the Gaussian factor. We now
can consider the polynomial excitations of this ground state and moreover
the behind integrability of this system.

\subsection{From ground state to the Hamiltonian}

As we have done in the case of Calogero-Sutherland model, we now propose
a Hamiltonian exactly has this Laughlin state as its ground state,
that is,
\begin{eqnarray}
H_{L} & = & \sum_{i}(\partial_{i}+\partial_{i}(\ln\tilde{\Psi}\{z_{i}\}))(\partial_{i}-\partial_{i} (\ln\tilde{\Psi}\{z_{i}\})).
\end{eqnarray}
Here $\partial_{i}=\partial/\partial x_{i}\equiv z_{i}\partial_{z_{i}}$,
or equivalently, $z_{i}=e^{x_{i}}$. By separating out the ground
state contribution as follows, we have

\begin{eqnarray}
\tilde{{H}} & = & \tilde{(\Psi}\{z_{i}\})^{-1}H_{L}\tilde{\Psi}\{z_{i}\}\nonumber \\
 & = & \sum_{i}\left(z_{i}\partial_{z_{i}}+2z_{i}\partial_{z_{i}}(\ln\tilde{\Psi}\{z_{i}\})z_{i}\partial_{z_{i}}\right)\label{eq:Htilde}\\
 & = & \sum_{i}(z_{i}\partial_{z_{i}})^{2}-2b^{2}\sum_{i<j}\frac{z_{i}^{2}\partial_{z_{i}}}{z_{i}-z_{j}}\nonumber \\
 & = & \sum_{i}(z_{i}\partial_{z_{i}})^{2}-b^{2}\sum_{i<j}\frac{z_{i}^{2}\partial_{z_{i}}-z_{j}^{2}\partial_{z_{j}}}{z_{i}-z_{j}}\,.\nonumber
\end{eqnarray}
When acting on the normal ordered vertex operators, we have
\begin{eqnarray}
z_{i}\partial_{z_{i}}\prod_{j}^{N}\exp\left(b\sum_{n,j}\frac{a_{-n}}{n}z_{j}^{n}\right)|0\rangle & = & b\sum_{n}a_{-n}z_{i}^{n}\prod_{j}^{N}\exp\left(b\sum_{n,j}\frac{a_{-n}}{n}z_{j}^{n}\right)|0\rangle\,,\nonumber \\
(z_{i}\partial_{z_{i}})^{2}\prod_{j}^{N}\exp\left(b\sum_{n,j}\frac{a_{-n}}{n}z_{j}^{n}\right)|0\rangle & = & b^{2}\sum_{n,m}a_{-n}a_{-m}z_{i}^{n+m}\prod_{j}^{N}\exp\left(b\sum_{n,j}\frac{a_{-n}}{n}z_{j}^{n}\right)|0\rangle\nonumber \\
 & + & b\sum_{n}na_{-n}z_{i}^{n}\prod_{j}^{N}\exp\left(b\sum_{n,j}\frac{a_{-n}}{n}z_{j}^{n}\right)|0\rangle\,,\nonumber \\
a_{k}\prod_{j}^{N}\exp\left(b\sum_{n,j}\frac{a_{-n}}{n}z_{j}^{n}\right)|0\rangle & = & \sum_{i}bz_{i}^{n}\prod_{j}^{N}\exp\left(b\sum_{n,j}\frac{a_{-n}}{n}z_{j}^{n}\right)|0\rangle\,,\label{eq:coherent}\\
\frac{z_{i}^{2}\partial_{z_{i}}-z_{j}^{2}\partial_{z_{j}}}{z_{i}-z_{j}}\prod_{j}^{N}\exp\left(b\sum_{n,j}\frac{a_{-n}}{n}z_{j}^{n}\right)|0\rangle\nonumber \\
=\frac{b\sum_{n}a_{-n}\left(z_{i}^{n+1}-z_{j}^{n+1}\right)}{z_{i}-z_{j}} &  & \prod_{j}^{N}\exp\left(b\sum_{n,j}\frac{a_{-n}}{n}z_{j}^{n}\right)|0\rangle\nonumber \\
=b\sum_{n}a_{-n}(z_{i}^{n}+z_{i}^{n-1}z_{j}+ & \cdots & +z_{i}z_{j}^{n-1}+z_{j}^{n})\prod_{j}^{N}\exp\left(b\sum_{n,j}\frac{a_{-n}}{n}z_{j}^{n}\right)|0\rangle\,,\label{eq:interactionterm}
\end{eqnarray}
and substituting ($\ref{eq:coherent}$) into ($\ref{eq:Htilde}$),
we have the free part of ($\ref{eq:Htilde}$) as
\[
\tilde{{H}_{0}}=\sum_{n>0}na_{-n}a_{n}+b\sum_{n,m>0}a_{-n}a_{-m}a_{n+m}
\]
We need to be careful with the interaction term ($\ref{eq:interactionterm}$).
The summation on $i,\, j$ is not arbitrary. To turn this
term into bosonic operator formalism, we have to do a trick as follow.
Firstly, we change the summation from $i<j$ to $i\neq j$. It
gives rise to a simple $\frac{1}{2}$ factor. Secondly, we add $i=j$ terms into
the summation and then finally subtract these terms.
Follow these steps, we have
\begin{eqnarray}
b^{3}\sum_{i<j}\,\sum_{n>0}a_{-n}(z_{i}^{n}+z_{i}^{n-1}z_{j}+&\cdots&+z_{i}z_{j}^{n-1}+z_{j}^{n})  =  \frac{b}{2}\sum_{n,m>0}a_{-n-m}a_{n}a_{m}\nonumber \\
-\frac{1}{2}b^{2}\sum_{n>0}na_{-n}a_{n} & + & b^{2}(N-1)\sum_{n>0}a_{-n}a_{n\,,}\label{eq:interoper}
\end{eqnarray}
The last term of ($\ref{eq:interoper}$) is an irrelevant total energy
of free quasi-particles. When acting on a level $n$ state, it
gives rise to an eigen-energy
\[
E_{ir}|n\rangle=b^{2}(N-1)n|n\rangle\,
\]
as expected which describes a system of $N-1$ copies of non-interacting
bosonic oscillators with the frequency $b^{2}$. Therefore we ignore
its contribution and now we obtain the operator formalism of $H_{L}$

\begin{equation}
H_{L}=\sum_{n}\left(1-\frac{b^{2}}{2}\right)na_{-n}a_{n}+\sum_{n,m}b\left(a_{-n}a_{-m}a_{n+m}+\frac{1}{2}a_{-n-m}a_{n}a_{m}\right)\,.\label{eq:HLaughlin}
\end{equation}

\subsection{Fermionization of $H_{L}$}

Define the deformed bosonic modes as
\begin{eqnarray}
2b^{-1}\tilde{a}_{n} & = & a_{n}\, , \quad \frac{1}{2}b\tilde{a}_{-n}=a_{-n\,,}\label{eq:deformLau}
\end{eqnarray}
so that the triple-$a$ terms in ($\ref{eq:HLaughlin}$) now becomes
\begin{eqnarray*}
&&\sum_{n,m>0}\left(\frac{1}{2}b^{2}\tilde{a}_{-n}\tilde{a}_{-m}\tilde{a}_{n+m}+\tilde{a}_{-n-m}\tilde{a}_{n}\tilde{a}_{m}\right)\\
&=&\sum_{n,m>0}\left(\tilde{a}_{-n}\tilde{a}_{-m}\tilde{a}_{n+m}+\tilde{a}_{-n-m}\tilde{a}_{n}\tilde{a}_{m}\right)  -  (1-\frac{b^{2}}{2})\tilde{a}_{-n}\tilde{a}_{-m}\tilde{a}_{n+m}\,.
\end{eqnarray*}
It is straightforward to write down the fermionic formalism as the case of
the Calogero-Sutherland model in previous section. Hence we get
\begin{equation}
H_{L}=H_{0}+(1-\frac{b^{2}}{2})H_{d}+(1-\frac{b^{2}}{2})H_{t}\equiv H_{l}+(1-\frac{b^{2}}{2})H_{t}\,.\label{eq:LaughlinHamilton}
\end{equation}
The eigen-energy for this Hamiltonian is
\begin{equation}
E_{\lambda}^{Lau}=\sum_{i=1}^{\lambda_{1}^{t}}(\lambda_{i})^{2}-\frac{b^{2}}{2}\sum_{i=1}^{\lambda_{1}}(\lambda_{i}^{t})^{2}\,.\label{eq:eigenLau}
\end{equation}
The corresponding excitation state is
\[
|P_{\lambda}^{Lau}\rangle=D^{Lau}\frac{1}{1-\frac{1-b^{2}/2}{E_{\lambda}^{Lau}-H_{l}}H_{t}}s_{\lambda}(\psi\psi^{*})|vac\rangle\,,
\]
where $D^{Lau}$ is the similarity transformation related to ($\ref{eq:deformLau}$).
\begin{equation}
D^{Lau}=\exp\bigl(-\log(b/2)(\tilde{q}\tilde{a}_{0}+\sum_{n>0}\tilde{a}_{-n}\tilde{a}_{n}/n)\bigr)\,,
\end{equation}
and the fermionic expression:
\begin{equation}
D_{\pm}^{Lau}=\left(\frac{b}{2}\right)^{\pm\sum_{r>0}(\psi_{-r}\psi_{r}^{*}+\psi_{-r}^{*}\psi_{r})}\,.
\end{equation}
Notice that the polynomial is not a new polynomial, it is a Jack polynomial
with the parameter $\beta=\frac{b^{2}}{2}\,.$

Since the integrability of the Laughlin theory is exactly the same as that
of CS model we ignore its analysis here.

\section{Halperin State and Two-layer System}

Now we can consider the two-layer system. The corresponding ground
state wavefunction is the Halperin state, which reads
\begin{equation}
\tilde{\Psi}_{H}\left({z_{i}},{w_{j}}\right)=\prod_{i<j}^{N}(z_{i}-z_{j})^{p}\prod_{m<n}^{M}(w_{m}-w_{n})^{q}\prod_{i,m}^{N,M}(z_{i}-w_{m})^{r}\,.\label{eq:Halperin}
\end{equation}
The complexity of this wave-function lies in that it involves an interaction
between two layers. Let us first define the bosonic fields
\begin{eqnarray*}
\phi^{1}(z) & = & q_{0}^{1}+a_{0}^{1}\ln z+\sum_{n\neq0}\frac{a_{-n}^{1}}{n}z^{n}\,,\\
\phi^{2}(w) & = & q_{0}^{2}+a_{0}^{2}\ln w+\sum_{n\neq0}\frac{a_{-n}^{2}}{n}w^{n}\,,
\end{eqnarray*}
with commutation relation
\[
[a_{n}^{I}\,,\, a_{m}^{J}]=n\delta^{IJ}\delta_{n+m,0}\,,\, n,m\in\mathbb{Z},\, I,J\in{1,2}\,.
\]
The related coordinate system is defined as
\begin{eqnarray*}
Z_{I} & = & z_{I},\,\mbox{if}\,\ I\leq N\,,\\
Z_{I} & = & w_{I-N}\,,\, if\ \, N<I\leq N+M\,,\\
\partial_{I} & = & Z_{I}\partial_{Z_{I}}\,,
\end{eqnarray*}
we can now write down the Hamiltonian of this system as
\begin{eqnarray*}
H^{Hal} & = & \sum_{I}(\partial_{I}-2\partial_{I}(\ln\tilde{\Psi}_{H}(Z_{I}))\partial_{I}\\
 & = & \sum_{i=1}^{N}\left(z_{i}\partial_{z_{i}}-2z_{i}\partial_{z_{i}}\left(\ln\prod_{i<j}(z_{i}-z_{j})^{p}\right)\right.\\
 &  & +\left.2z_{i}\partial_{z_{i}}\left(\ln\prod_{i,m}(z_{i}-w_{m})^{r}\right)\right)z_{i}\partial_{z_{i}}\\
 & + & \sum_{m=1}^{M}\left(w_{m}\partial_{w_{m}}-2w_{m}\partial_{w_{m}}\left(\ln\prod_{m<n}(w_{m}-w_{n})^{q}\right)\right.\\
 &  & +\left.2w_{m}\partial_{w_{m}}\left(\ln\prod_{i,m}(z_{i}-w_{m})^{r}\right)\right)w_{m}\partial_{w_{m}}\\
 & = & H_{L}(p)+H_{L}(q)+2r\sum_{m,i}\frac{z_{i}^{2}\partial_{z_{i}}}{z_{i}-w_{m}}-2r\sum_{m,i}\frac{w_{m}^{2}\partial_{w_{m}}}{z_{i}-w_{m}}\,,\\
 & = & H_{L}(p)+H_{L}(q)+2r\sum_{n\geq0,m,i}\left(\frac{w_{m}}{z_{i}}\right)^{n}z_{i}\partial_{z_{i}}-2r\sum_{n>0,m,i}\left(\frac{w_{m}}{z_{i}}\right)^{n}w_{m}\partial_{w_{m}}\,\,,
\end{eqnarray*}
where $H_{L}(p)$ is the Laughlin Hamiltonian defined before in
which $b^{2}\rightarrow p\,.$ The vertex operators in this system
are
\[
V_{\sqrt{p}}^{1}(z)=e^{\sqrt{p}\phi^{1}(z)}\,,\: V_{\sqrt{q}}^{2}(w)=e^{\sqrt{q}\phi^{2}(w)}\,.
\]
The differential operators relate to operators and power-sum polynomials
are defined as follows
\begin{eqnarray*}
z_{i}\partial_{z_{i}}\prod_{i=1}^{N}V_{\sqrt{p}}^{1,-}(z_{i}) & = & \sum_{n>0}\sqrt{p}a_{-n}^{1}z_{i}^{n}\prod_{i=1}^{N}V_{\sqrt{p}}^{1,-}(z_{i})\,,\\
w_{m}\partial_{w_{m}}\prod_{j=1}^{M}V_{\sqrt{q}}^{2,-}(w_{j}) & = & \sum_{n>0}\sqrt{q}a_{-n}^{2}w_{j}^{n}\prod_{n=1}^{M}V_{\sqrt{q}}^{2,-}(w_{j})\,,\\
a_{n}^{1}\prod_{i=1}^{N}V_{\sqrt{p}}^{1,-}(z_{i})|0\rangle & = & \sqrt{p}\sum_{i}z_{i}^{n}\prod_{i=1}^{N}V_{\sqrt{p}}^{1,-}(z_{i})|0\rangle\\
a_{n}^{2}\prod_{j=1}^{N}V_{\sqrt{p}}^{2,-}(w_{j})|0\rangle & = & \sqrt{q}\sum_{i}w_{j}^{n}\prod_{i=1}^{N}V_{\sqrt{p}}^{1,-}(w_{j})|0\rangle
\end{eqnarray*}
Therefore we have
\begin{eqnarray*}
2r\sum_{n\geq0,m,i}\left(\frac{w_{m}}{z_{i}}\right)^{n}z_{i}\partial_{z_{i}} & \rightarrow & 2r\frac{1}{\sqrt{q}}\sum_{\begin{array}[t]{c}
n\geq0\\
m>0
\end{array}}a_{-n}^{1}a_{n-m}^{1}a_{m}^{2}\,,\\
2r\sum_{n>0,m,i}\left(\frac{w_{m}}{z_{i}}\right)^{n}w_{m}\partial_{w_{m}} & \rightarrow & 2r\frac{1}{\sqrt{p}}\sum_{\begin{array}[t]{c}
n>0\\
m>0
\end{array}}a_{-m}^{1}a_{-n}^{2}a_{n+m}^{2}\,.
\end{eqnarray*}
It is now easy to write down the bosonic operator formalism, we have
\begin{eqnarray*}
H^{Hal} & = & H_{L}(p,a^{1})+H_{L}(q,a^{2})+H_{int}\,,\\
H_{int} & = & 2r\left(\frac{1}{\sqrt{q}}\sum_{\begin{array}[t]{c}
n\geq0\\
m>0
\end{array}}a_{-n}^{1}a_{n-m}^{1}a_{m}^{2}-\frac{1}{\sqrt{p}}\sum_{\begin{array}[t]{c}
n>0\\
m>0
\end{array}}a_{-m}^{1}a_{-n}^{2}a_{n+m}^{2}\right)\\
 & = & r\left(\frac{1}{\sqrt{q}}\sum_{m>0}L_{-m}^{1}a_{m}^{2}+\sum_{n,m>0}(\frac{1}{\sqrt{q}}a_{-n}^{1}-\frac{2}{\sqrt{p}}a_{-n}^{2})a_{-m}^{1}a_{n+m}^{2}\right)\,.
\end{eqnarray*}
Notice that the $H_{int}$ is a fascinating interaction. It is always a triangular term such that it
subtracts boxes in Young diagram $\mu$ on layer-2 and adds the same
number of boxes into Young diagram $\lambda$ on layer-1. The Hilbert space of this Hamiltonian is
expanded by coupled bi-Jack functions, that is,
\[
|\Omega_{\lambda,\mu}^{0}\rangle=|P_{\lambda}^{1}(p/2)\rangle\otimes|P_{\mu}^{2}(q/2)\rangle\,.
\]
It is not an eigenstate of the Halperin Hamiltonian, but only a highest
weight state of the eigenstate, which can be obtained as following
formula
\[
|\Omega_{\lambda,\mu}^{r}\rangle=\frac{1}{1-\frac{1}{E_{\lambda,\mu}-H_{L}(p)-H_{L}(q)}H_{int}}|\Omega_{\lambda,\mu}^{0}\rangle\,,
\]
where the eigen-energy is written as
\[
E_{\lambda,\mu}=\sum_{i}^{\lambda_{1}^{t}}(\lambda_{i})^{2}+\sum_{j}^{\mu_{1}^{t}}(\mu_{j})^{2}-\frac{p}{2}\sum_{k}^{\lambda_{1}}(\lambda_{k}^{t})^{2}-\frac{q}{2}\sum_{l}^{\mu_{1}}(\mu_{k}^{t})^{2}\,.
\]
The fermionization of the Hamiltonian as in ($\ref{eq:Halperin}$)
is not hard. Let us first consider the first term in $H_{int}$ . It
is
\begin{eqnarray*}
H_{int}^{1} & = & \frac{2r}{\sqrt{q}}\sum_{\begin{array}[t]{c}
n\geq0\\
m>0
\end{array}}a_{-n}^{1}a_{n-m}^{1}a_{m}^{2}\\
 & = & \frac{2r}{\sqrt{q}}\left(\sum_{\begin{array}[t]{c}
r,s,k,l,u\in\mathbb{Z}+\frac{1}{2}\\
r+s+k+l=-m<0
\end{array}}:\psi_{r}^{1}\psi_{s}^{1*}\psi_{k}^{1}\psi_{l}^{1*}\psi_{-u}^{2}\psi_{m+u}^{2*}:\right.\\
 &  & \hspace{1em}\hspace{1em}+\left.\sum_{u\in\mathbb{Z}+\frac{1}{2},m>0}:\psi_{-u}^{2}\psi_{m+u}^{2*}:(\text{Contractions})\right)\,,
\end{eqnarray*}
where the contractions is calculated as
\begin{eqnarray*}
\text{Contractions} & = & \sum_{\begin{array}[t]{c}
\begin{array}[t]{c}
n,r>0\end{array}\end{array}}:(\psi_{-r}^{1}\psi_{-n+r}^{1*}-\psi_{-r-n}^{1*}\psi_{r}^{1})(\psi_{-s}^{1}\psi_{n-m+s}^{1*}-\psi_{n-m-s}^{1*}\psi_{s}^{1}):\,,\\
 & = & \sum_{r>0}(\psi_{-r}^{1}\psi_{r-m}^{1*}+\psi_{-m-s}^{1*}\psi_{s}^{1})\,.
\end{eqnarray*}
Therefore
\begin{eqnarray*}
H_{int}^{1} & = & \frac{2r}{\sqrt{q}}\left(\sum_{\begin{array}[t]{c}
r,s,k,l,u\in\mathbb{Z}+\frac{1}{2}\\
r+s+k+l=-m<0
\end{array}}:\psi_{r}^{1}\psi_{s}^{1*}\psi_{k}^{1}\psi_{l}^{1*}\psi_{-u}^{2}\psi_{m+u}^{2*}:\right.\\
 &  & \hspace{1em}\hspace{1em}+\left.\sum_{\begin{array}[t]{c}
u>0\\
r,m>0
\end{array}}(\psi_{-u}^{2}\psi_{m+u}^{2*}-\psi_{m-u}^{2*}\psi_{u}^{2})(\psi_{-r}^{1}\psi_{r-m}^{1*}+\psi_{-m-r}^{1*}\psi_{r}^{1})\right)\,.
\end{eqnarray*}
Similarly, we have
\begin{eqnarray*}
H_{int}^{2} & = & -\frac{2r}{\sqrt{p}}\sum_{\begin{array}[t]{c}
n,m>0\end{array}}a_{-m}^{1}a_{-n}^{2}a_{n+m}^{2}\\
 & = & -\frac{2r}{\sqrt{p}}\left(\sum_{\begin{array}[t]{c}
r,s,k,l,u\in\mathbb{Z}+\frac{1}{2}\\
r+s+k+l=m>0
\end{array}}:\psi_{r}^{2}\psi_{s}^{2*}\psi_{k}^{2}\psi_{l}^{2*}\psi_{-u}^{1}\psi_{-m+u}^{1*}:\right.\\
 &  & \hspace{1em}\hspace{1em}+\left.\sum_{u\in\mathbb{Z}+\frac{1}{2},m>0}:\psi_{-u}^{1}\psi_{-m+u}^{1*}:(\text{Contractions})\right)\\
 &  & =-\frac{2r}{\sqrt{p}}\left(\sum_{\begin{array}[t]{c}
r,s,k,l,u\in\mathbb{Z}+\frac{1}{2}\\
r+s+k+l=m>0
\end{array}}:\psi_{r}^{2}\psi_{s}^{2*}\psi_{k}^{2}\psi_{l}^{2*}\psi_{-u}^{1}\psi_{-m+u}^{1*}:\right.\\
 &  & \hspace{1em}\hspace{1em}+\left.\sum_{\begin{array}[t]{c}
u,m>0\\
r>0
\end{array}}(\psi_{-u}^{1}\psi_{-m+u}^{1*}-\psi_{-m-u}^{1*}\psi_{u}^{1})(\psi_{-r}^{2}\psi_{r+m}^{2*}+\psi_{m-r}^{2*}\psi_{r}^{2})\right)\,.
\end{eqnarray*}
Though this fermionic formalism is not effective in calculating the
eigenstate, it plays an important role in deriving the behind tau-function
and its Hirota integrability of this system. Actually, the fermionic
formalism completely defines the fermionic orbit of a generator of
$GL(\infty)$ which in turn determine the tau-function of this theory.
We are working in detail in this direction.

\subsection{Similarity transformation}

As explained before, we need to do a similarity transformation to
recover the deformed operator formalism of eigenstates to a standard
operator formalism. It is easy to write down this similarity transformation,
that is
\[
D^{Hal}=D^{Lau}(b^{2}\rightarrow p,\, a\rightarrow a^{1})D^{Lau}(b^{2}\rightarrow q,\, a\rightarrow a^{2})\,.
\]

\section{Conclusions and Future Works}

In conclusion, we introduce a systematic way to extract the
integrability of several models. For CS model, we express the Hamiltonian
in bosonic as well as fermionic representations. The eigenstate and
eigenvalue are obtained explicitly. The construction of Jack state,
in the fermionic representation, is highly involved in the fermionic
triangularization of fermionic Hamiltonian of the CS model. For Laughlin
and Halperin states, which can be seen as solitonic wavefunction,
we construct the corresponding Hamiltonians in the same manner as
CS model. We obtain their bosonic and fermionic representations.
The explicit solutions, e.g. excitations and eigenvalues, are
exactly solved. The integrability of Laughlin state, is the same as
that of CS model, while for Halperin state, the integrability is determined
also by the triangularization nature of the Hamiltonian.

There are several problems worthy of exploring in the future. Firstly,
though the integrability in this article are inherited from free fermions,
it is not clear to the authors that how to construct the integrable
hierarchies. In soliton theory, the integrable hierarchy can be
determined by the Lax operators. The Lax method is not the expected
one for solving the problem since there are in general integration
operators (the pseudo-differential operators) additional to usual differential
operators. A possible solution may be the inverse scattering method,
which will relate the integrable hierarchy to the inverse scattering
equation\citep{das1989integrable}. This hierarchy tells us how integral
of motions can be constructed by a recursive relation. A higher level
integral of motion determines a refiner excitation structure of
the model\citep{jimbo1983solitons}. Secondly, for FQHEs, people believe
special Jack polynomial may be related to certain FQHE wavefunction.
It still remains mysterious to us what kind of constraint leads
to a truncation of the fusion rule of Jack polynomials. Thirdly, we
expect a direct generalization of our analysis to Haldane-Shastry model\citep{haldane1988exact,shastry1988exact,talstra1995integrals},
or the spin CS model, we are working on that.

\section*{Acknowledgement}
The authors are grateful to Morningside Center of Chinese Academy of Sciences and Kavli Institute for Theoretical Physics China at the Chinese Academy of Sciences for providing excellent research environment and financial support to our seminar in mathematical physics. Currently Jie Yang is a visiting scholar in the Department of Physics at University of California, Berkeley and is grateful for the hospitality during the 2014-2015 academic year. The project is partially supported by National Natural Science Foundation of China (No. 11401400), Returned Oversea Students Fund in Beijing, and Specialized Research Fund for the Doctoral Program of Higher Education (No. 20121108120005).
\section*{Appendix}

We will denote a partition by its parts $(\lambda_1, \lambda_2, \cdots, \lambda_{\lambda^t_1})$ and the Frobenius notation $(\alpha_1, \cdots, \alpha_d|\beta_1, \cdots, \beta_d) $ as well where $d$ is the diagonal length of $\lambda$. With this notation $n_i$ and $m_i$ in theorem 4 are related with $\alpha_i$ and $\beta_i$ in such a way that $n_i=\alpha_i+1/2$ and $m_i=\beta_i+1/2$.
Theorem 4 is then written as
\begin{eqnarray*}
 \sum_{i=1}^{\lambda_1}(\lambda^t_i)^2&=& \sum_{i=1}^d\left[\frac{\alpha_i}{3}+\beta_i^2+\frac{2\beta_i}{3}\right]\\
&&\quad +\frac{2}{3}d\left[\sum_{i=1}^d(\alpha_i+2\beta_i+\frac{3}{2})
+\sum_{i=1}^d2\alpha_i(i-1)\right.\\&&\left.-\sum_{i=1}^d(d-i)\alpha_i+\sum_{i=1}^d\beta_i(i-1)-2\sum_{i=1}^d\beta_i(d-i)\right]\\
&=&\sum_{i=1}^d(\beta_i^2+2i\beta_i+2i\alpha_i-\alpha_i)+d^2.
\end{eqnarray*}
 To prove this theorem we need two preliminary steps.

 Step 1:
\begin{equation}\label{eq:kappa}
2[n(\lambda^t)-n(\lambda)]=\sum_{i=1}^d \alpha_i(\alpha_i+1)-\beta_i(\beta_i+1).
\end{equation}
It is obvious to prove the first step by using two very useful identities among several multi-number sets, namely
\begin{equation*}
  \{\beta_i, (i\leq d)\}=\{0, 1, \cdots, \lambda_1^t-1\}-\{i-\lambda_i-1, (d+1\leq i\leq \lambda_1^t) \},
\end{equation*}
and similarly
\begin{equation*}
  \{\alpha_i, (i\leq d)\}=\{0, 1, \cdots, \lambda_1-1\}-\{i-\lambda^t_i-1, (d+1\leq i\leq \lambda_1)\}.
\end{equation*}
These two identities are also very useful in proving step 2. Now let us compute
\begin{eqnarray*}
  2[n(\lambda^t)-n(\lambda)]&=&\sum_{i=1}^{\lambda_1^t} \lambda_i(\lambda_i-2i+1)\\
  &=& \sum_{i=1}^{\lambda_1^t} (\lambda_i-i)^2+|\lambda|-\sum_{i=1}^{\lambda_1^t}i^2\\
  &=& \sum_{i=1}^d \alpha_i^2 +\sum_{i=d+1}^{\lambda_1^t} (-\lambda_i+i-1+1)^2 +|\lambda|
-\sum_{i=1}^{\lambda_1^t}i^2.
\end{eqnarray*}
We apply one of the two identities. Therefore one term becomes
\begin{equation*}
  \sum_{i=d+1}^{\lambda_1^t} (-\lambda_i+i-1+1)^2 =\sum_{i=0}^{\lambda^t_1-1}(i+1)^2-\sum_{i=1}^d (\beta_i+1)^2
\end{equation*}
and since
\begin{equation*}
|\lambda|=\sum_{i=1}^d(\alpha_i+\beta_i+1),
\end{equation*}
combining all terms together we  obtain the result in (\ref{eq:kappa}).

Step 2:

\begin{equation}\label{eq:hook}
2[n(\lambda^t)+n(\lambda)]=\sum_{i=1}^d (\alpha_i^2+\beta_i^2+4 i\alpha_i+4 i\beta_i -3\alpha_i-3\beta_i)+2d(d-1).
\end{equation}

\begin{figure}
  \centering
  \includegraphics[width=0.5\textwidth]{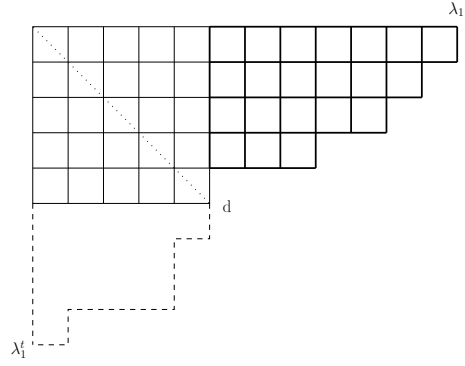}
  \caption{The three regions of a Young diagram are separated by solid lines, thick lines, and dashed lines. }
  \label{fig:regions}
\end{figure}

To prove (\ref{eq:hook}), we recall a formula in Macdonald's book \cite{macdonald1995symmetric}
\begin{equation*}
  \sum_{x\in\lambda}h(x)=\sum_{(i, j)\in\lambda}(\lambda_i-i+\lambda^t_j-j+1)=n(\lambda^t)+n(\lambda)+|\lambda|.
\end{equation*}
Hence
\begin{equation*}
  2[n(\lambda^t)+n(\lambda)]=2\sum_{(i, j)\in\lambda} (\lambda_i -i +\lambda^t_j-j).
\end{equation*}
Now let us compute
\begin{eqnarray*}
  \sum_{(i, j)\in\lambda}\lambda_i-i+\lambda_j^t-j&=&\sum_{i=1}^{\lambda^t_1}\sum_{j=1}^{\lambda_i} \lambda_i-i+\lambda_j^t-j \\
&=& \left[\sum_{i=1}^d\sum_{j=1}^d+\sum_{i=1}^d\sum_{j=d+1}^{\lambda_i}+\sum_{j=1}^d\sum_{i=d+1}^{\lambda^t_j}\right] (\lambda_i-i+\lambda_j^t-j).
\end{eqnarray*}
The regions of the summation are shown in Fig. \ref{fig:regions}.
The square region surrounded by solid lines is simply
\begin{equation*}
  \sum_{i=1}^d\sum_{j=1}^d (\lambda_i-i+\lambda_j^t-j)
=d\sum_{i=1}^d (\alpha_i+\beta_i).
\end{equation*}
Now let us compute the sum for the region surrounded by the thick lines.
\begin{equation*}
  \sum_{i=1}^d\sum_{j=d+1}^{\lambda_i}(\lambda_i-i)=\sum_{i=1}^d(\lambda_i-i)(\lambda_i-d)=\sum_{i=1}^d(\lambda_i^2-i\lambda_i-d\lambda_i+id).
\end{equation*}
We obtain
\begin{equation*}
  \sum_{i=1}^d\sum_{j=d+1}^{\lambda_i}(\lambda_j^t-j+1-1)
=-\sum_{i=1}^d\left[\sum_{j=0}^{\lambda_i-i} j-(\sum_{j=i}^d\alpha_j)\right]-\sum_{i=1}^d(\lambda_i-d)
\end{equation*}
where we have used
\begin{equation*}
  \{j-\lambda_j^t-1, (d+1\leq j\leq \lambda_i)\}=\{0, 1, 2, \cdots, \lambda_i-i\}-\{\alpha_j, (i\leq j\leq d)\}.
\end{equation*}
Therefore
\begin{eqnarray*}
    \sum_{i=1}^d\sum_{j=d+1}^{\lambda_i}(\lambda_j^t-j+1-1)&=&-\sum_{i=1}^d\frac{(\lambda_i-i)(\lambda_i-i+1)}{2}+\sum_{i=1}^d i\alpha_i-\sum_{i=1}^d(\lambda_i-d)\\
&=&\sum_{i=1}^d\left[-\frac{\lambda_i^2}{2}+i\lambda_i-\frac{i(i-1)}{2}-\frac{\lambda_i}{2}+i\alpha_i-(\lambda_i-i+i-d)\right].
\end{eqnarray*}
Combining them we get
\begin{eqnarray*}
  \sum_{i=1}^d\sum_{j=d+1}^{\lambda_i} (\lambda_i-i+\lambda_j^t-j) &=&\sum_{i=1}^d \left[\frac{(\lambda_i-i)^2}{2}+i(\lambda_i-i)+i\alpha_i-d(\lambda_i-i)-\frac{3}{2}(\lambda_i-i)-i+d\right]\\
&=&\sum_{i=1}^d\left[\frac{\alpha_i^2}{2}+2i\alpha_i-d\alpha_i-\frac{3\alpha_i}{2}+d-i \right].
\end{eqnarray*}
Similarly we can compute the region surrounded by the dashed lines and the result is just replacing $\alpha$ with $\beta$ in above formula.
Hence we have
\begin{equation*}
  \sum_{(i, j)\in\lambda} \lambda_i -i +\lambda^t_j-j= \sum_{i=1}^d\left[ \frac{\alpha_i^2}{2}+\frac{\beta_i^2}{2}+2i \alpha_i+2i\beta_i-\frac{3\alpha_i}{2}-\frac{3\beta_i}{2}+2(d-i)\right]
\end{equation*}
Therefore twice of it will give rise to (\ref{eq:hook}).

Now let us compute
\begin{equation*}
  4n(\lambda)=2[n(\lambda^t)+n(\lambda)]-2[n(\lambda^t)-n(\lambda)]=\sum_{i=1}^d(2\beta_i^2+4i\alpha_i+4i\beta_i-4\alpha_i-2\beta_i)+2d(d-1).
\end{equation*}
Since
\begin{equation*}
4n(\lambda)=2\sum_{i=1}^{\lambda_1}\lambda^t_i(\lambda^t_i-1)=2\sum_{i=1}^{\lambda_1}(\lambda^t_i)^2-2\sum_{i=1}^d\alpha_i-2\sum_{i=1}^d\beta_i-2d,
\end{equation*}
We obtain
\begin{equation*}
  \sum_{i=1}^{\lambda_1}(\lambda^t_i)^2=\sum_{i=1}^d(\beta_i^2+2i\alpha_i-\alpha_i+2i\beta_i )+d^2.
\end{equation*}

\end{document}